\documentclass[12pt,english,journal=jacsat]{achemso}
\setkeys{acs}{etalmode=truncate, maxauthors=10}
\usepackage[T1]{fontenc}
\usepackage[latin9]{inputenc}
\usepackage{color}
\usepackage{amsbsy}
\usepackage{amstext}
\usepackage{graphicx}
\usepackage{subscript}
\usepackage{setspace}

\makeatletter

\title{More bridging ligands activate direct exchange: the case of anisotropic Kitaev effective magnetic interactions}

\author{Pritam Bhattacharyya}
\affiliation{Institute for Theoretical Solid State Physics, Leibniz IFW Dresden, Helmholtzstra{\ss}e~20, 01069 Dresden, Germany}
\email{pritambhattacharyya01@gmail.com}

\author{Nikolay A. Bogdanov}
\affiliation{Max Planck Institute for Solid State Research, Heisenbergstra{\ss}e 1, 70569 Stuttgart, Germany}

\author{Liviu Hozoi}
\affiliation{Institute for Theoretical Solid State Physics, Leibniz IFW Dresden, Helmholtzstra{\ss}e~20, 01069 Dresden, Germany}
\email{l.hozoi@ifw-dresden.de}

\@ifundefined{showcaptionsetup}{}{%
 \PassOptionsToPackage{caption=false}{subfig}}
\usepackage{subfig}
\makeatother

\usepackage{babel}

\begin{document}

\singlespacing

\begin{abstract}
\noindent
A magnet is a collection of magnetic moments.
How those interact is determined by what lies in between.
In transition-metal and rare-earth magnetic compounds, the configuration of the ligands around each
magnetic center and the connectivity of the ligand cages are therefore pivotal --- for example, the
mutual interaction of magnetic species connected through one single ligand is qualitatively different
from the case of two bridging anions.
Two bridging ligands are encountered in Kitaev magnets.
The latter represent one of the revelations of the 21st century in magnetism research:
they feature highly anisotropic intersite couplings with seemingly counterintuitive directional
dependence for adjacent pairs of magnetic sites and unique quantum spin-liquid ground states that
can be described analytically.
Current scenarios for the occurrence of pair-dependent magnetic interactions as proposed by Kitaev
rely on {\it indirect} exchange mechanisms based on intersite electron hopping.
Analyzing the wavefunctions of Kitaev magnetic bonds at both single- and multi-configuration levels,
we find however that {\it direct}, Coulomb exchange may be at least as important,
in 5$d$ and 4$d$ $t_{2g}^5$, 3$d$ $t_{2g}^5e_g^2$, and even rare-earth 4$f^1$ Kitaev-Heisenberg
magnets.
Our study provides concept clarification in Kitaev magnetism research and the essential reference
points for reliable computational investigation of how novel magnetic ground states can be engineered
in Kitaev, Kitaev-Heisenberg, and Heisenberg edge-sharing systems.
\end{abstract}


\maketitle


\noindent
{\it \ --- \ ToC Graphic \ --- \ }

\

\begin{figure}[]
\includegraphics[width=0.7\textwidth]{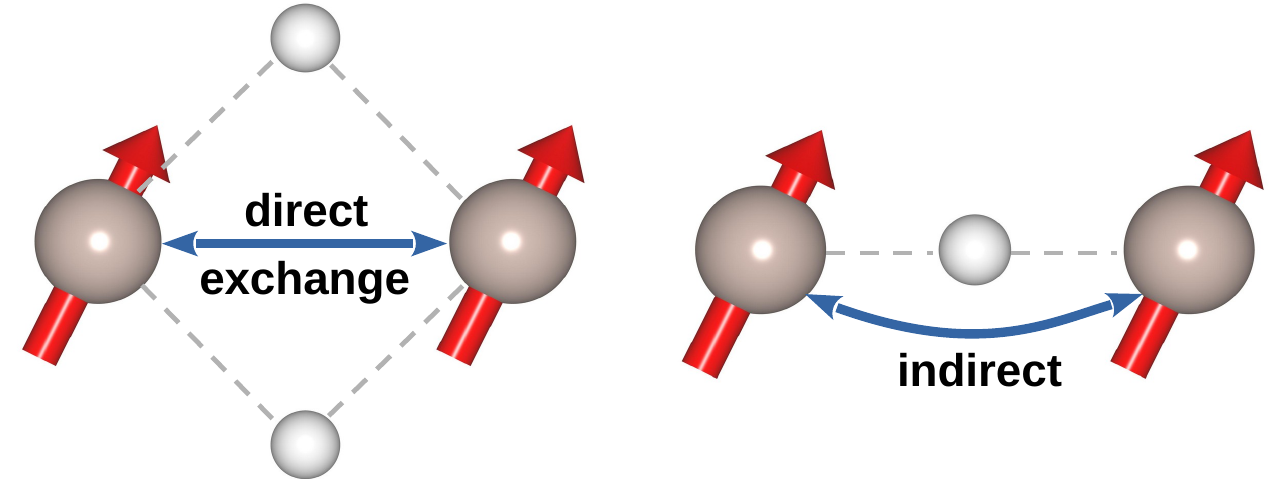}
\end{figure}

\

\noindent
Electronic-level magnetism rests on the notion of exchange.
{\it Direct} exchange occurs through the interplay of Pauli's exclusion principle and Coulomb repulsion,
as discussed by Heisenberg, Dirac, and van Vleck already in the 1920s \cite{Heisenberg1926,Dirac1926,
VanVleck-32}, has no classical analogue, and is the main effect responsible for ferromagnetism.
The antiferromagnetic ground states observed in a variety of magnetic systems, on the other hand, 
arise from {\it indirect} exchange interactions involving intersite electron hopping: M-M kinetic
exchange, where only electrons at the magnetic centers (Ms) are active, and M-L-M superexchange, where
electrons at nonmagnetic, intermediary ionic sites [e.\,g., chalcogenide or halide ligands (Ls)] are
also swapped.

In phenomenological effective interaction models with one, half-filled orbital per magnetic ion and
180$^{\circ}$ M-L-M chemical bonds, kinetic exchange and superexchange imply rather simple analytical
expressions.
Such physics took center stage in studies of copper oxide compounds, e.\,g., cuprate superconductors
\cite{Eskes_1993}, leaving direct exchange in the shade.
More recently, kinetic exchange and superexchange were discussed in the context of anisotropic Kitaev
interactions \cite{Kitaev2006} on networks of edge-sharing $t_{2g}^5$, \cite{Khaliullin_2005,
Khaliullin_PRL_2009,Takagi2019} $t_{2g}^5e_g^2$, \cite{Liu_et_al} and 4$f^1$ ML$_6$
octahedra \cite{Motome_2020}.
%
However, different from the case of corner-sharing ML$_6$ octahedra and 180$^{\circ}$ M-L-M links
in superconducting cuprates (where the direct M-M orbital overlap is small), for edge-sharing ML$_6$
units and 90$^{\circ}$ (or $\approx$90$^{\circ}$) M-L-M paths \cite{Khaliullin_PRL_2009,
Takagi2019,Browne,Kanyolo}
direct exchange may in principle become comparable in size with the indirect exchange mechanisms,
especially for M-site orbitals with lobes along the M-M axis.
Yet, direct exchange has been completely ignored so far in phenomenological Kitaev-Heisenberg exchange
models \cite{Khaliullin_PRL_2009,Takagi2019,Khaliullin_2005,Liu_et_al,Motome_2020}.

Even for corner-sharing ML$_n$ units, there are situations where direct exchange may again compete
with the indirect, hopping-mediated exchange mechanisms: strongly bent M-L-M paths, especially in the
cases of adjacent pyramidal ML$_5$ entities, adjacent ML$_4$ tetrahedra, and mixed types of polyhedra,
e.\,g., networks of corner-sharing ML$_6$ octahedra and ML$_4$ tetrahedra.
Mingled polyhedra --- in particular, octahedra and tetrahedra --- are encountered in some of the most
promising multiferroic/magnetoelectric materials, i.e., the Y-type hexaferrites \cite{Kocsis2019}, 
and in the family of Fe$_2$Mo$_3$O$_8$ \cite{Ghara2023} and Co$_2$Mo$_3$O$_8$ \cite{Szaller_et_al} 
multiferroics.

How direct and indirect exchanges work in the case of known Kitaev-Heisenberg systems is illustrated
at {\it ab initio} level in the following, by means of wavefunction electronic-structure theory
{\color{red} \cite{olsen_bible,fulde_book}.}


\noindent
{\it The A$_3$BM$_2$L$_6$ material platform, $t_{2g}^5$ vs $t_{2g}^5e_g^2$ Kitaev centers.\,}
Anisotropic Kitaev intersite interactions may occur on both honeycomb and triangular networks
of edge-sharing ML$_6$ octahedra and are characterized by peculiar directional dependence of
the leading
anisotropic coupling $K\tilde{S}_{i}^{\gamma}\tilde{S}_{j}^{\gamma}$\,:\, 
for a given pair of adjacent 1/2 pseudospins ${\mathbf{\tilde{S}}}_{i}$ and ${\bf{\tilde{S}}}_{j}$,
the easy axis defined through the index $\gamma$ can be parallel to either $x$, $y$, or $z$ \cite{
Kitaev2006,Khaliullin_PRL_2009}.
Triangular networks of edge-sharing ML$_6$ units are encountered in e.\,g.~rhombohedral crystalline
structures derived from the rocksalt setting, in the form of successive sheets perpendicular to the
111 direction (see, e.\,g., discussion in ref.~\cite{Browne}).
Hexagonal architectures can be obtained out of the triangular layers if certain magnetic sites are
removed or occupied by nonmagnetic atomic species \cite{Browne,Kanyolo,Khaliullin_PRL_2009}.

Many triangular and hexagonal magnets can be generically described through the chemical formula
A$_3$BM$_2$L$_6$ (sometimes written as A$_3$M$_2$BL$_6$) \cite{Kanyolo}.
For example\,:\,
B can be Li in the spin-liquid honeycomb iridate H$_3$LiIr$_2$O$_6$ \cite{Kitagawa2018} or Sb in the
cobaltates Li$_3$Co$_2$SbO$_6$ \cite{Brown_et_al} and Na$_3$Co$_2$SbO$_6$ \cite{NCSO_wong_16};
A=B=Na, M=Ir, and L=O gives Na$_2$IrO$_3$, a representative 5$d$ Kitaev-Heisenberg honeycomb magnet
\cite{Takagi2019};
A=B=0 (i.e., empty A and B sites), M=Ru, and L=Cl corresponds to RuCl$_3$, a 4$d$ Kitaev-Heisenberg
honeycomb system \cite{Takagi2019};
with B=M we arrive to AML$_2$ delafossite-type triangular structures, e.\,g., NaRuO$_2$ \cite{Ortiz2023,
Bhattacharyya2023}, CoI$_2$ (with unoccupied A sites) \cite{Kim2023}, and RbCeO$_2$ \cite{Ortiz_PRM};
A=B=M=Co and L=O corresponds to rocksalt CoO (i.e., successive triangular Co-ion and O layers normal
to the 111 axis).

For the case of edge-sharing ML$_6$ octahedra with $t_{2g}^5$ valence electron configuration at the
magnetic sites, the interplay of $t_{2g}$-shell spin-orbit coupling, intersite hopping, and on-site
(Hund) exchange were shown to generate anisotropic exchange \`a la Kitaev \cite{Kitaev2006} ({\it
indirect}, hopping mediated) already 15 years ago \cite{Khaliullin_PRL_2009}.
However, the {\it direct}, Coulomb M-M exchange

\begin{figure}[]
\centering
\includegraphics[width=0.7\linewidth]{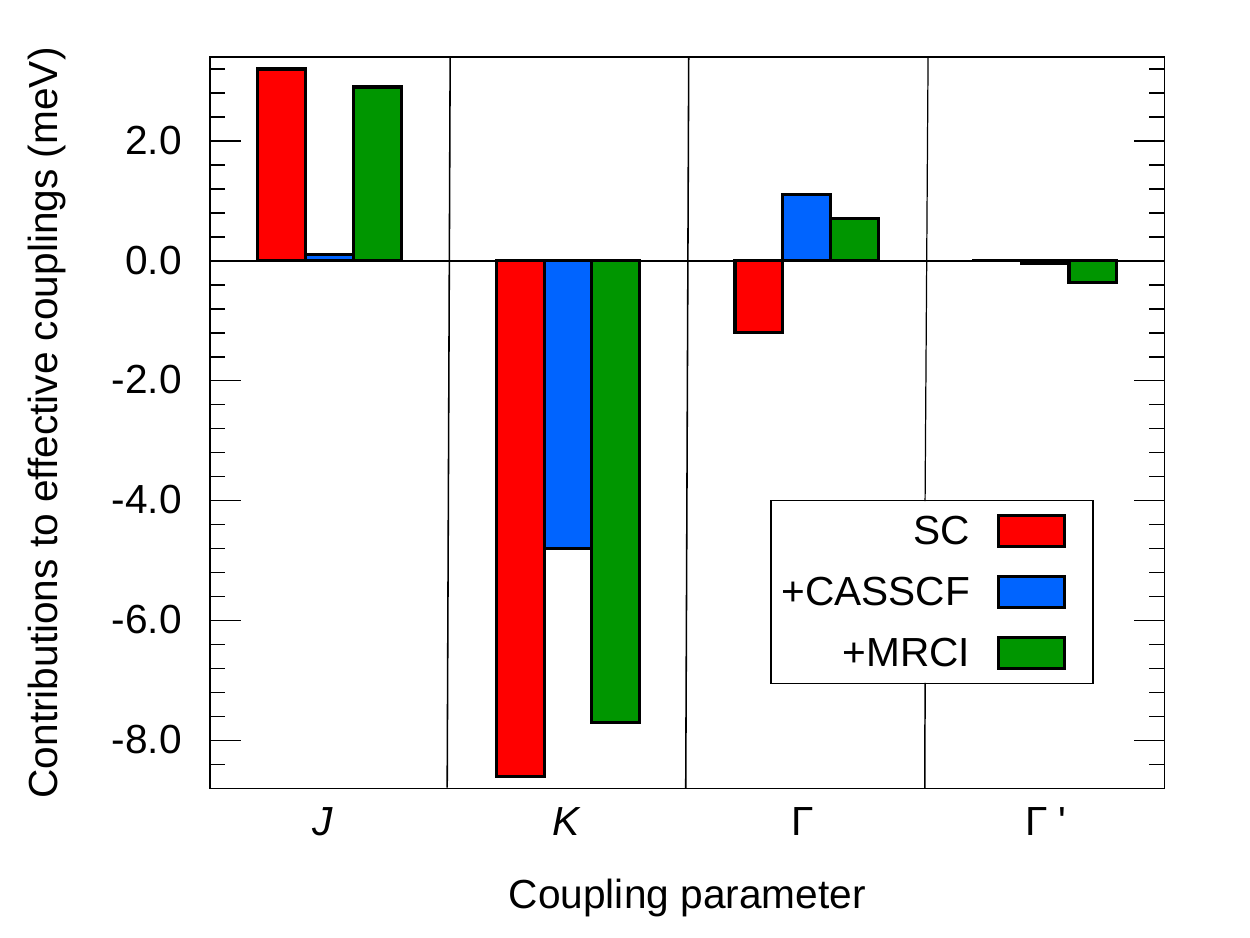}
\caption{
Exchange contributions to the intersite magnetic couplings in 5$d^5$ Na$_2$IrO$_3$:
Coulomb exchange (SC results, in red),
Ir($t_{2g}$)--Ir($t_{2g}$) kinetic exchange (as the difference between CASSCF and SC data, in blue),
plus contributions related to Ir-O$_2$-Ir superexchange, Ir($t_{2g}$)$\rightarrow$Ir($e_g$) excitations,
and so called dynamical correlation effects \cite{olsen_bible}
(as the difference between MRCI and CASSCF, in green).
}
\label{Ir213}
\end{figure}

\noindent
amplitudes should also be sizable, especially those
implying $\sigma$- and $\pi$-type pairs of orbitals --- 
the interplay between such orbital-dependent Coulomb exchange and $t_{2g}$-shell spin-orbit coupling
is another possible source of anisotropic magnetism.
The roles of the different mechanisms can be easily verified with {\it ab initio} wavefunction
electronic-structure computational methods \cite{olsen_bible}.
Such calculations have been used for a long time to explore solid-state electronic
structures and can provide information that is not accessible by other means, on e.\,g.~nontrivial
correlated wavefunctions
\cite{Petersen2023,Petersen2022}, cohesive energies \cite{Yang_et_al}, band gaps \cite{Stoyanova_et_al,
Vo_et_al}, and, of particular interest here, exchange mechanisms \cite{Martin_1993,VANOOSTEN1996,
Bogdanov2022}.

Focusing first on the hitherto neglected direct exchange mechanism, spin-orbit calculations that
account only for the leading $t_{2g}^5$-$t_{2g}^5$ ground-state electron configuration (to which we
refer as single-configuration, SC, computations) and subsequent mapping \cite{Yadav2016} onto the 
effective nearest-neighbor spin Hamiltonian (see Supporting Information for further details)
indicate indeed large contributions. 
Those are shown as red bars in Fig.\;1 and Fig.\;2, for Na$_2$IrO$_3$ and RuCl$_3$, prototype $t_{2g}^5$
Kitaev-Heisenberg honeycomb magnets \cite{Khaliullin_PRL_2009,Takagi2019}. 
%
Besides the isotropic Heisenberg $J$ and diagonal anisotropic $K$ couplings, the off-diagonal
$\Gamma$ and $\Gamma'$ effective coupling parameters are analyzed as well in the two figures.
They enter the effective Hamiltonian for a pair of adjacent 1/2-pseudospins ${\bf{\tilde{S}}}_i$ and
${\bf{\tilde{S}}}_j$ as \cite{Takagi2019} 

\begin{equation}
\label{eqn:Hamil}
\mathcal{H}_{ij}^{(\gamma)} = J{\bf{\tilde{S}}}_i\cdot {\bf{\tilde{S}}}_j + \\
                              K\tilde{S}_i^{\gamma}\tilde{S}_j^{\gamma} + \\
                              \sum_{\alpha\neq\beta} \Gamma_{\alpha\beta} \\
                              (\tilde{S}_i^{\alpha}\tilde{S}_j^{\beta} +  \\
                              \tilde{S}_i^{\beta}\tilde{S}_j^{\alpha}) \ ,
\end{equation}
with $\alpha,\beta,\gamma\!\in\!\{x,y,z\}$.
For e.\,g.~a $z$-type M-M bond (i.\,e., M$_2$L$_2$ plaquette normal to the $z$ axis),
$\Gamma\!\equiv\!\Gamma_{xy}$ and $\Gamma'\!\equiv\!\Gamma_{yz}\!=\!\Gamma_{zx}$.

\begin{figure}[]
\centering
\includegraphics[width=0.7\linewidth]{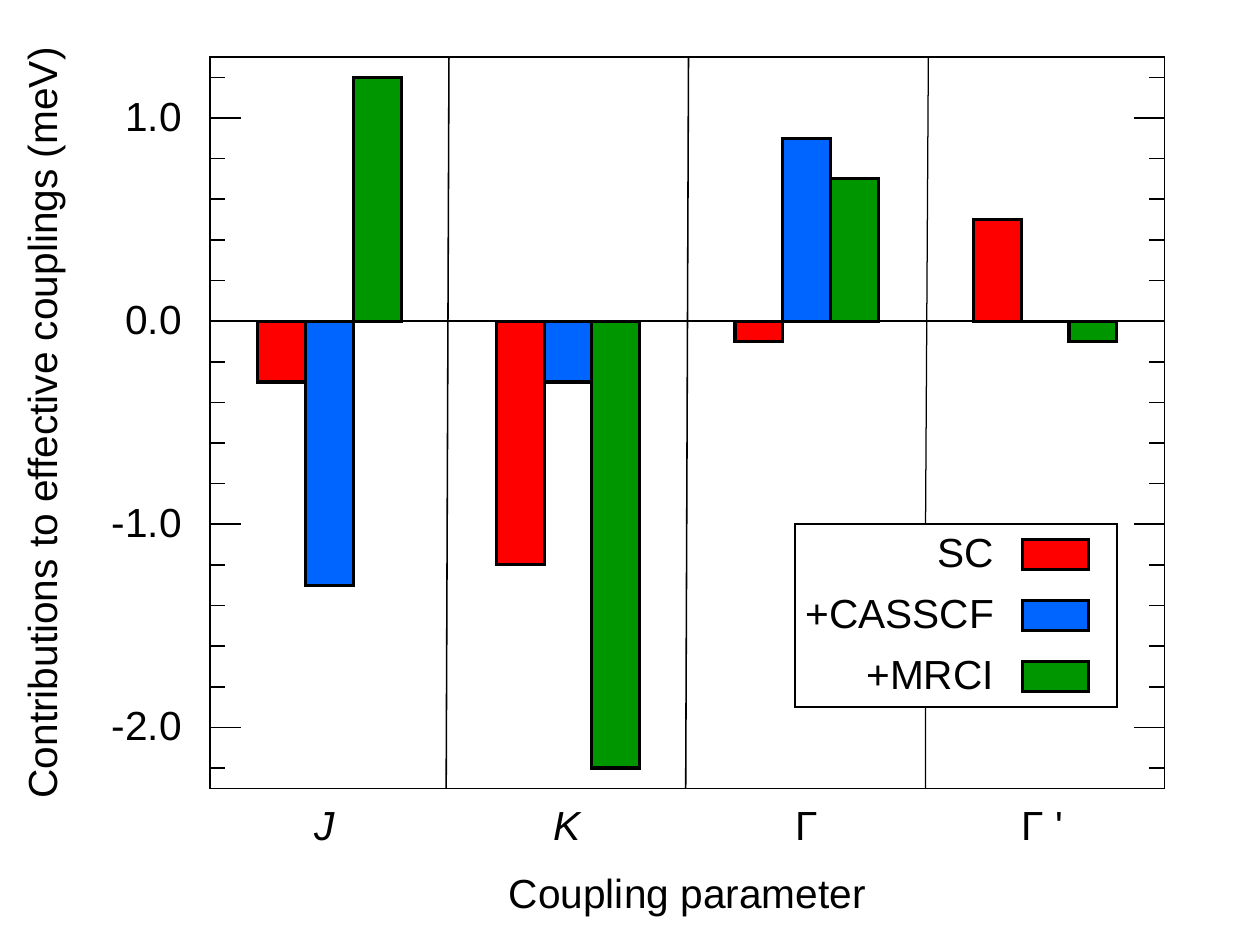}
\caption{
Contributions to the intersite magnetic couplings in 4$d^5$ RuCl$_3$:
Coulomb exchange (red bars),
Ru($t_{2g}$)--Ru($t_{2g}$) kinetic exchange (blue),
plus contributions related to Ru-Cl$_2$-Ru superexchange, Ru($t_{2g}$)$\rightarrow$Ru($e_g$) excitations,
and dynamical correlation (green).
}
\label{Ru013}
\end{figure}

The indirect mechanisms, kinetic exchange (blue) and superexchange (green), require more involved
computations, multiconfiguration complete-active-space self-consistent-field (CASSCF) wavefunction
expansions \cite{olsen_bible,MCSCF_Molpro} that account for intersite excitations within the
transition-ion $t_{2g}$ sector and  multireference configuration-interaction (MRCI) wavefunctions
\cite{olsen_bible,MRCI_Molpro} including also L-to-M excitations, respectively (see Supporting
Information for computational details).
%
Remarkably, direct exchange brings the largest contributions to $K$, $J$, and $\Gamma$ in 5$d$
Na$_2$IrO$_3$ (as shown in Fig.\;1) and to $\Gamma'$ in 4$d$ RuCl$_3$ (Fig.\;2).
It also provides sizable weight to the Kitaev coupling $K$ in RuCl$_3$, $\approx$33\%.

The role of direct exchange is even more spectacular in the case of $t_{2g}^5e_g^2$\, 3$d$ compounds,
e.\,g., Li$_3$Co$_2$SbO$_6$:
direct exchange is the dominant exchange mechanism for all four effective parameters, as illustrated
in Fig.\,3.
To clearly identify the role of kinetic exchange, two
different sets of multiconfiguration
calculations were performed: first accounting only for on-site intra-3$d$ excitations, referred to as
single-site complete-active-space (SSCAS, with contributions depicted in light blue in Fig.\,3) and
then for all possible intra-3$d$ excitations, both on-site and intersite (with additional contributions
shown in darker blue).
The numerical values obtained at different levels of approximation are provided in Table \ref{tab:exchange}.

\noindent
{\it 4$f^1$-4$f^1$ anisotropic direct exchange.\,} 
Recently quantum spin liquid (QSL) behavior has been reported in a number of triangular-lattice
pseudospin-1/2 4$f^{13}$ and 4$f^1$ chalcogenides:
YbMgGaO$_4$ \cite{Shen2016}, NaYbS$_2$ \cite{Baenitz2018}, NaYbO$_2$ \cite{
Bordelon2019}, NaYbSe$_2$ \cite{Dai}, CsYbSe$_2$ \cite{Pai}, KYbSe$_2$ \cite{Xing}, RbYbSe$_2$ \cite{
Xing}, and RbCeO$_2$ \cite{Ortiz_PRM}.
Given the smaller (or comparable \cite{Ortiz_PRM}) energy scale of the 4$f$ crystal-field splittings
with respect to the strength of the spin-orbit coupling $\lambda$, there are 7\,$\times$\,7\,=\,49
configurations

\begin{figure}[]
\centering
\includegraphics[width=0.7\linewidth]{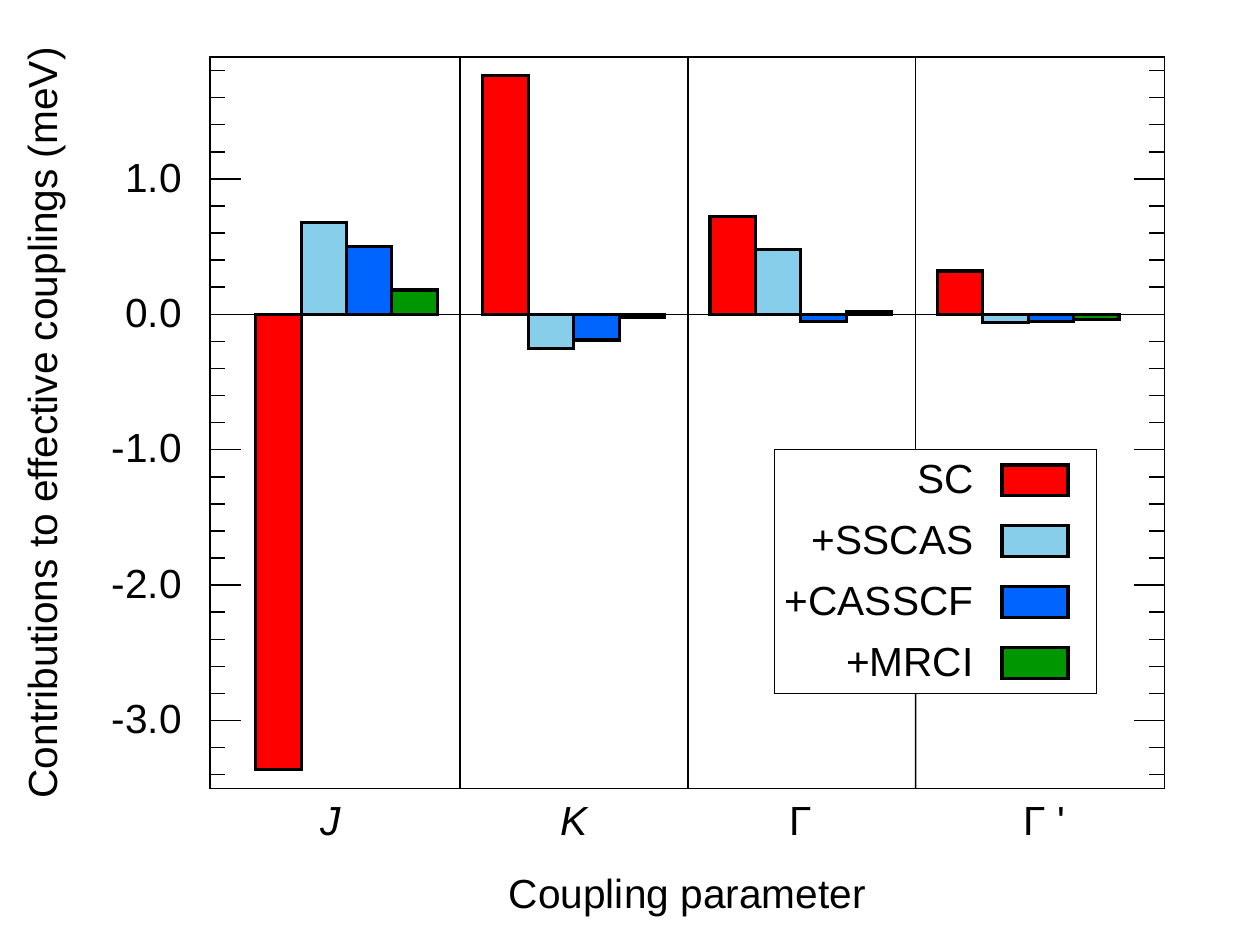}
\caption{
Exchange mechanisms contributing to the intersite magnetic couplings in 3$d^7$ Li$_3$Co$_2$SbO$_6$:
$d$-$d$ Coulomb exchange (red bars),
renormalization due to on-site intra-3$d$ excitations (light blue),
$d$-$d$ kinetic exchange (dark blue),
plus contributions related to Co-O$_2$-Co superexchange and dynamical correlation (green).
}
\label{lcso}
\end{figure}

\begin{table}[t]
\caption{
Effective magnetic couplings at different levels of approximation for $C_{2h}$ M$_2$L$_{10}$
two-octahedra units in the Kitaev-Heisenberg systems Li$_3$Co$_2$SbO$_6$ \cite{Brown_et_al},
RuCl$_3$ \cite{Cao_et_al}, Na$_2$IrO$_3$ \cite{Choi_et_al}, and RbCeO$_2$ \cite{Ortiz_PRM}.
%
%
}
\begin{tabular}{l  r  r  r  r }
\hline
\hline\\[-0.20cm]

                  &$J$
                            &$K$
                                   &$\Gamma$
                                               &$\Gamma^\prime$   \\

\hline\\[-0.15cm]
3$d$ Li$_3$Co$_2$SbO$_6$ (meV)\\

SC                &--3.4    &1.8   &0.7        &0.3  \\
SSCAS             &--2.7    &1.5   &1.2        &0.3  \\
CASSCF            &--2.2    &1.3   &1.1        &0.2  \\
MRCI              &--2.0    &1.3   &1.2        &0.2  \\

                \\[-0.15cm]
4$d$ RuCl$_3$ (meV)  \\

SC                &--0.3    &--1.2   &--0.1      &0.5  \\
CASSCF            &--1.6    &--1.5   &0.8        &0.5  \\
MRCI              &--0.4    &--3.7   &1.5        &0.4  \\

		    \\[-0.15cm]
5$d$ Na$_2$IrO$_3$ (meV) \\

SC                &3.2    &--8.6    &--1.2      &0.01    \\
CASSCF            &3.3    &--13.4   &--0.1      &--0.04\\
MRCI              &6.2    &--21.1   &0.6        &--0.4     \\

        	\\[-0.15cm]
4$f$ RbCeO$_2$ ($\mu$eV) \\

SSCAS           &--10.3    &--37.3    &--9.1      &--7.1\\
CASSCF          &59.4    &--28.3    &--8.8      &--5.4\\
\hline
\hline
\end{tabular}
\label{tab:exchange}
\end{table}

\noindent
that must be explicitly considered in the spin-orbit treatment for 4$f^1$-4$f^1$  and
4$f^{13}$-4$f^{13}$ pairs of ions.
The single-site ground-state Kramers doublet is typically separated from the lowest on-site excitations
by a sizable gap; when mapping the {\it ab initio} data onto the effective two-site magnetic model,
considering only the lowest four `magnetic' states out of the whole set of 196 is then a good
approximation.
The model-Hamiltonian studies on 4$f^1$-4$f^1$  and 4$f^{13}$-4$f^{13}$ (super)exchange are also
performed along this idea \cite{Mironov_1996,Onoda,Rau,Motome_2020}.

Mapping the lowest four eigenstates obtained by spin-orbit 4$f$ SSCAS and 4$f$ CASSCF
two-octahedra computations
onto the effective magnetic Hamiltonian described by (1), it was possible
to estimate the role of direct and kinetic exchange, respectively, for the effective intersite couplings
in 4$f^1$ RbCeO$_2$ (Fig.\;4), a triangular-lattice rare-earth system that does not order magnetically
down to 60 mK \cite{Ortiz_PRM}.
It is found that for the anisotropic channel ($K$, $\Gamma$, and $\Gamma'$) the direct exchange
contributions are very important (see also the data in Table \ref{tab:exchange}), larger than what kinetic exchange
brings.
Spin-orbit MRCI computations for two adjacent CeO$_6$ octahedra (to estimate Ce-O$_2$-Ce superexchange
contributions) are computationally quite demanding and will constitute the topic of a different
study.


\noindent
{\it Discussion.\,}
A 21st-century revelation in magnetism research is Kitaev's honeycomb-lattice anisotropic spin model,
in particular, the seemingly counterintuitive directional dependence of its anisotropic intersite
couplings, the peculiar flavor of QSL ground state that the model hosts, and the possibility of
describing the QSL analytically \cite{Kitaev2006}.
With Khaliullin's and Jackeli's remarkable intuition and pioneering work \cite{Khaliullin_PRL_2009,
Khaliullin_2005}, we know how anisotropic (pseudo)spin interactions \`a la Kitaev may arise in quantum
matter and in which kind of magnets we should look for those.
However, it appears that the Kitaev (pseudo)spin interaction tableau is not yet fully uncovered:
through {\it ab initio}, wavefunction computations here we reveal an additional Kitaev interaction
mechanism --- direct, Coulomb exchange (also referred to as potential exchange) in the presence of
sizable spin-orbit coupling.
It turns out that in prototype Kitaev-Heisenberg magnets such as Na$_2$IrO$_3$ and Li$_3$Co$_2$SbO$_6$
it actually represents the leading intersite interaction.
Moreover, it seemingly brings important contributions to the anisotropic interactions on 4$f$-ion
triangular lattices.

The massive Coulomb exchange contributions reported here represent very solid data, all those are
obtained at the lowest possible level of approximation in {\it ab initio} electronic-structure theory,
Hartree-Fock-like.
Similar results on the magnitude of the intersite Coulomb exchange contributions should be obtained by
density-functional computations using functionals that build in exact (i.e., Hartree-Fock) exchange and
completely disregard correlations\footnote{
On the other hand, describing kinetic exchange and superexchange through the exchange-correlation functional
remains elusive.
}.

Direct, Coulomb exchange adds a new dimension to the Kitaev-Heisenberg interaction landscape.
An important aspect that needs to be understood is the interplay of direct and indirect exchange
mechanisms, e.\,g., how those different contributions can be tuned to 0 in the case of the Heisenberg
$J$, such that the Kitaev QSL phase is stabilized.
This would provide theoretical guidelines to, e.\,g., experiments under strain on Kitaev-Heisenberg
magnets.
That the different exchange mechanisms may compete with each other is apparent in Fig.\;2, for
the isotropic component in RuCl$_3$:
direct and kinetic exchange (red and blue

\begin{figure}[]
\centering
\includegraphics[width=0.7\linewidth]{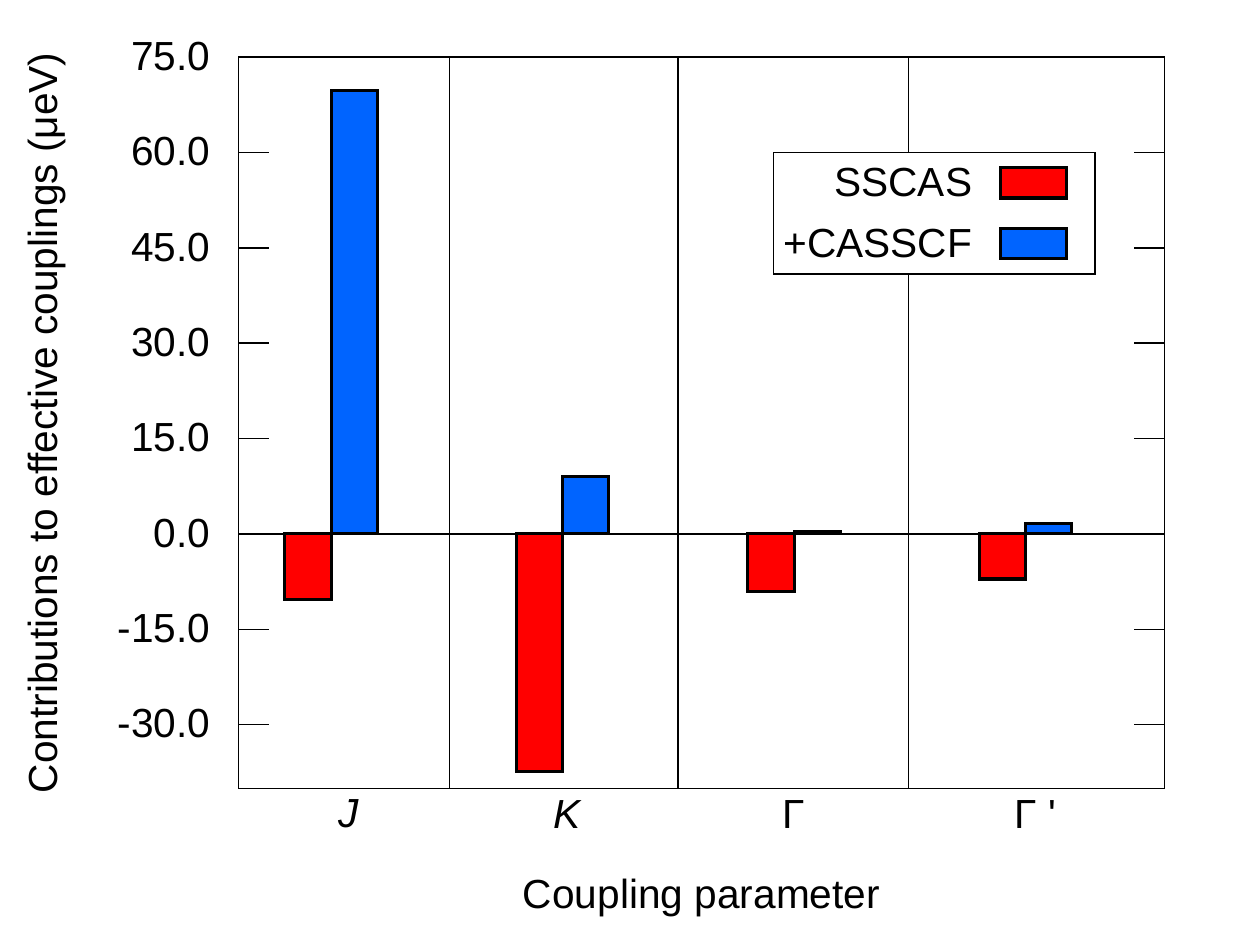}
\caption{
4$f$-4$f$ Coulomb exchange (red) and 4$f$-4$f$ kinetic exchange (blue) in RbCeO$_2$.
}
\label{rbceo2}
\end{figure}

\noindent
bars) compete with and are nearly counterbalanced by
superexchange and additional correlation effects accounted for in MRCI (green).
It is worth noting that the sum of the different effects in the isotropic channel agrees with
the small $J$ value derived from, e.\,g., neutron scattering measurements on  RuCl$_3$ \cite{RuCl3_ML}.
%
The analysis versus experimental data is also illustrative for the case of the A$_3$Co$_2$SbO$_6$
cobaltates: the {\it leading} Coulomb-exchange contribution --- ferromagnetic, isotropic, stronger in
Li$_3$Co$_2$SbO$_6$ (--3.4 meV, see Table\;I) than in Na$_3$Co$_2$SbO$_6$ (--1.4 meV \cite{NCSO_Pritam}) ---
seemingly explains
(i) the ferromagnetic Curie-Weiss temperatures found experimentally in these compounds
 \cite{NCSO_wong_16,McQueen_et_al} and
(ii) a Curie-Weiss temperature that is larger in Li$_3$Co$_2$SbO$_6$ (15 K \cite{McQueen_et_al}) than in
Na$_3$Co$_2$SbO$_6$ (2K \cite{NCSO_wong_16}).

\

\

\noindent
{\bf Competing Interests.\,}
The authors declare no competing interests.

\

\noindent
{\bf Data Availability.\,}
Raw quantum chemical data on which this manuscript is based will be made publicly available upon acceptance.

\

\noindent
{\bf Acknowledgments.}
We thank G.~Khaliullin, S.\;Nishimoto, U.\,K.\,R{\"o}{\ss}ler, D.\;Efremov, T.~Petersen, and
R.~C.~Morrow for discussions and U.~Nitzsche for technical support.
P.~B.~and L.~H.~acknowledge financial support from the German Research Foundation (Deutsche
Forschungsgemeinschaft, DFG), project number 468093414.

\

\noindent
{\bf Supporting Information.\,}
Detailed computational scheme, employed basis sets, and  orbital basis for computing exchange contributions
are discussed in the Supporting Information.


\

\bibliography{refs_jun11}
\end{document}


\title{Supporting Information --- More bridging ligands activate direct exchange\,:\,\\the case of anisotropic Kitaev
                                  effective magnetic interactions}

\author{Pritam Bhattacharyya}
\affiliation{Institute for Theoretical Solid State Physics, Leibniz IFW Dresden, Helmholtzstra{\ss}e~20, 01069 Dresden, Germany}

\author{Nikolay A. Bogdanov}
\affiliation{Max Planck Institute for Solid State Research, Heisenbergstra{\ss}e 1, 70569 Stuttgart, Germany}

\author{Liviu Hozoi}
\affiliation{Institute for Theoretical Solid State Physics, Leibniz IFW Dresden, Helmholtzstra{\ss}e~20, 01069 Dresden, Germany}

\date\today
\maketitle

\subsection{Methods}

\noindent
All quantum chemical computations were carried out using the {\sc molpro} suite of programs \cite{Molpro}.
For each type of embedded cluster, the crystalline environment was modeled as a large array of point
charges which reproduces the crystalline Madelung field within the cluster volume;
we employed the {\sc ewald} program \cite{Klintenberg_et_al} to generate the point-charge embeddings.

The many-body $ab$ $initio$ calculations were performed for fragments consisting of two central octahedra
and either four (for hexagonal lattices) or eight (for the triangular compound) adjacent octahedra.
CASSCF computations were carried out with six $t_{2g}$ orbitals and ten electrons as active for the
iridate and ruthenate systems, with the ten valence 3$d$ orbitals and 14 electrons in the active space
for the cobaltate, and with 14 4$f$ orbitals and two electrons for the 4$f^1$ system.
The CASSCF optimizations were performed for all possible spin multiplicities: lowest nine singlets
and nine triplets associated with the leading $t_{2g}^5$-$t_{2g}^5$ configuration for the iridate and
ruthenate, lowest nine singlet, nine triplet, nine quintet, and nine septet states associated with
the leading $t_{2g}^5e_g^2$-$t_{2g}^5e_g^2$ ground-state configuration for the cobaltate, and lowest
49 singlets and 49 triplets associated with the $f^1$-$f^1$ configuration for RbCeO$_2$.
Different from previous quantum chemical investigations (e.\,g., on RuCl$_3$ in ref.\;\cite{Yadav2016}),
where the core and semi-core orbitals were kept frozen at CASSCF level, as obtained from a preliminary
Hartree-Fock calculation preceding the CASSCF step, all orbitals were here reoptimized in the CASSCF
variational procedure.
Interestingly, for the particular case of RuCl$_3$, by full orbital optimization in CASSCF the sign
of the Heisenberg $J$ is reversed:
from $J\!=\!1.2$ meV in ref.\;\cite{Yadav2016}, we arrive at $J\!=-0.4$ meV in the final MRCI spin-orbit
computation (Table 1, main article) if all orbitals are reoptimized in CASSCF.
The other nearest-neighbor coupling parameters are less affected. 
In the subsequent MRCI correlation treatment, single and double excitations out of the central-unit
magnetic $d$/$f$ and bridging-ligand $p$ orbitals were considered (for the cobaltate, O 2$p_z$ only).
%
%
Spin-orbit couplings were further accounted for as described in \cite{SOC_Molpro}, either at SC, SSCAS,
CASSCF, or MRCI level.
The lowest four spin-orbit eigenstates from the {\sc molpro} output (with eigenvalues lower by $\sim$30
meV or more compared to other states) were mapped onto the eigenvectors
of the effective spin Hamiltonian 1 (see main article), following the procedure described in refs.~\cite{
Bogdanov2015,Yadav2016}.  

\begin{figure}[b]
\centering
\includegraphics[width=0.95\linewidth]{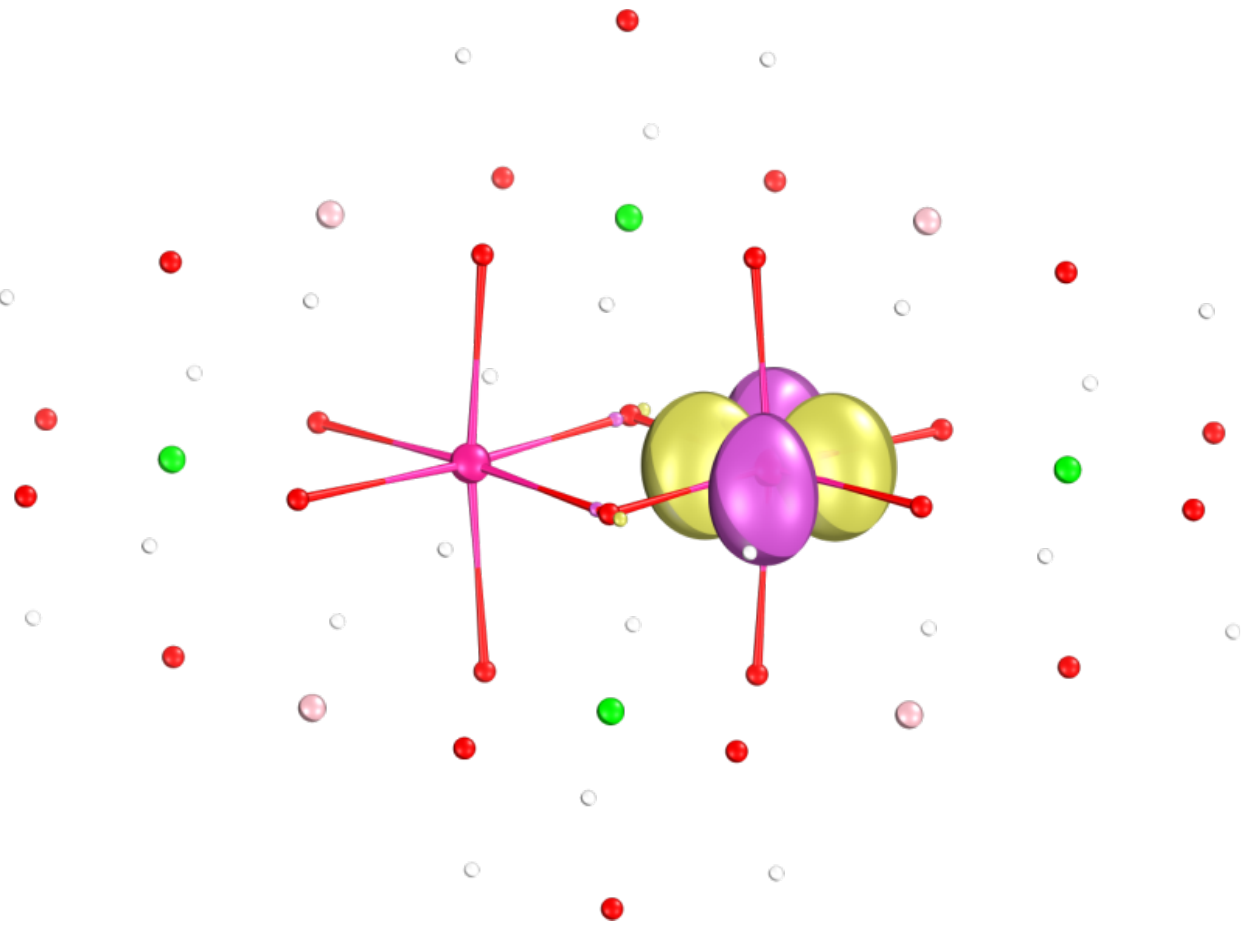}
\caption{
Localized Co 3$d$ $xy$ magnetic orbital in Li$_3$Co$_2$SbO$_6$, plot with 95\% of the electron density
within the contour; for plots with less than 94\% of the electron density within the contour, the O
$p$ tails are not at all visible.
Bonds are depicted only for the Co$_2$O$_{10}$ block of two edge-sharing octahedra;
other atomic sites shown in the figure define the quantum mechanical cluster described in the previous
section.
}
\label{Coxy}
\end{figure}

We used the Pipek-Mezey methodology \cite{PM_method} to obtain localized central-unit orbitals.
The localized orbitals (LOs) allow to construct SC wavefunctions (using appropriate restrictions in
the {\sc molpro} inputs for the occupations of the LOs) and subsequently derive the Coulomb exchange
contributions to the effective nearest-neighbor magnetic couplings (i.\,e., the red bars in Figs.\;1--4).
Illustrative LO plots and information concerning the atomic basis sets are provided in Supporting
Information.
Orbital composition analysis through Mulliken partition \cite{Lu_book,Lu_2024} yields 99\% Co 3$d$
character for the Co $t_{2g}$ LOs and 97\% Co 3$d$ character for the Co $e_g$ LOs in Li$_3$Co$_2$SbO$_6$,
94\% Ru 4$d$ character for the Ru $t_{2g}$ magnetic LOs in RuCl$_3$, 90\% Ir 5$d$ character for the Ir
$t_{2g}$ magnetic LOs in Na$_2$IrO$_3$, and 99.5\% Ce 4$f$ character for the magnetic LOs in RbCeO$_2$.
No orbital optimization was further performed in the SC and SSCAS computations;
the latter can be described as occupation-restricted multiple active space (ORMAS) CI calculations \cite{
ormas_ivanic_03}.

Lattice parameters as determined in \cite{Choi_et_al}, \cite{Cao_et_al}, \cite{Brown_et_al}, and
\cite{Ortiz_PRM} were respectively employed for Na$_2$IrO$_3$, $\alpha$-RuCl$_3$, Li$_3$Co$_2$SbO$_6$,
and RbCeO$_2$.

\subsection{Basis set information}

\noindent
{\it Na$_2$IrO$_3$.\,}
Relativistic pseudopotentials (ECP60MDF) and basis sets (BSs) of effective quadruple-$\zeta$ quality
(ECP60MDF-VTZ) \cite{Figgen_et_al} were utilized for the two `central' Ir ions.
All-electron BSs of quintuple-$\zeta$ quality were employed for the two bridging ligands \cite{O-ligands}
while all-electron triple-$\zeta$ BSs were applied for the remaining eight O anions
\cite{O-ligands} associated with the two octahedra of the reference magnetic unit.
The four adjacent transition ions were represented as closed-shell Pt$^{4+}$ $t_{2g}^6$ species,
using relativistic pseudopotentials (Ir ECP61MDF) and (Ir ECP60MDF-VDZ) (8s7p6d)/[3s3p3d] BSs \cite{
Figgen_et_al};
the $t_{2g}$ orbitals of these adjacent cations were part of the inactive orbital space.
The other 16 O ligands associated with the four adjacent transition metal sites were described through
minimal all-electron atomic natural orbital (ANO) BSs \cite{Pierloot1995}.
Large-core  pseudopotentials were employed for the 18 Na nearest neighbors \cite{Fuentealba_Na}.

\begin{figure}[b]
\centering
\includegraphics[width=0.95\linewidth]{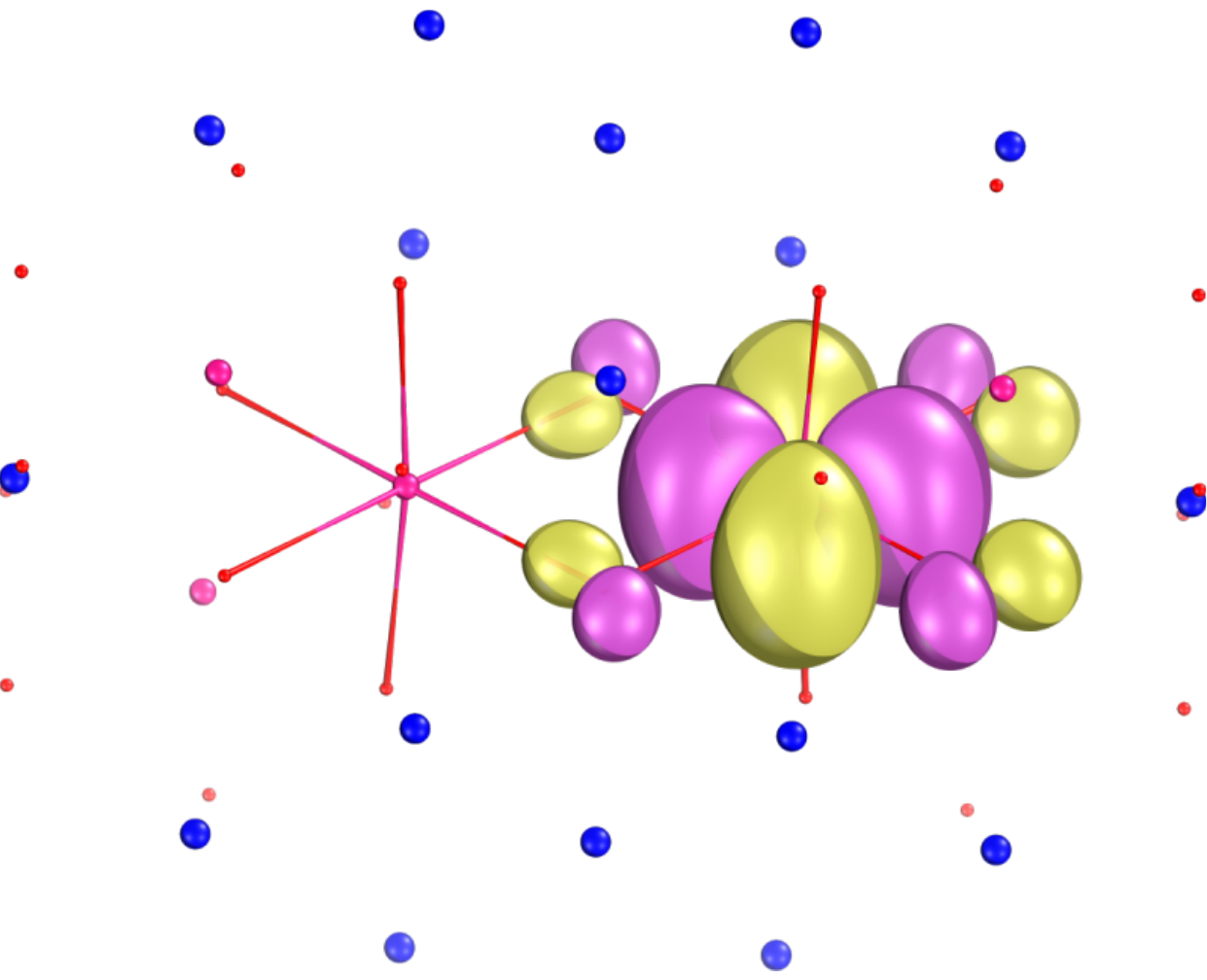}
\caption{
Localized Ir 5$d$ $xy$ magnetic orbital in Na$_2$IrO$_3$, plot with 90\% of the electron density
within the contour.
For comparison, a localized O 2$p$ valence orbital is depicted in Fig.\;S3.
}
\label{Irxy}
\end{figure}

\noindent
{\it $\alpha$-RuCl$_3$.\,}
We employed energy-consistent relativistic pseudopotentials (ECP28MDF) and Gaussian-type valence BSs
of effective quadruple-$\zeta$ quality (ECP28MDF-VTZ) \cite{4d_elements} for the central Ru species.
All-electron BSs of quintuple-$\zeta$ quality were utilized for the two bridging ligands \cite{
Dunning_Cl} and of triple-$\zeta$ quality for the remaining eight Cl anions \cite{Dunning_Cl} linked
to the two octahedra of the reference unit.
The four adjacent cations were represented as closed-shell Rh$^{3+}$ $t_{2g}^6$ species, using
relativistic pseudopotentials (Ru ECP29MDF) and (Ru ECP28MDF-VDZ) (8s7p6d)/[3s3p3d] BSs for electrons
in the 4th shell \cite{4d_elements};
the outer 16 Cl ligands associated with the four adjacent octahedra were described through minimal ANO
BSs \cite{Pierloot1995}.

\noindent
{\it Li$_3$Co$_2$SbO$_6$.\,}
We utilized all-electron BSs of quadruple-$\zeta$ quality for the central Co sites, [7s6p4d2f] \cite{
Co_basis}.
All-electron BSs of quintuple-$\zeta$ quality were employed for the two bridging ligands \cite{O-ligands}
while all-electron triple-$\zeta$ BSs were applied for the remaining eight O anions \cite{O-ligands}
associated with the two octahedra of the reference unit.
The four adjacent transition ions were represented as closed-shell Zn$^{2+}$ cations, using large-core
pseudopotentials Zn ECP28MWB plus uncontracted (3s2p) valence BSs \cite{Schautz1998}), and the four adjacent
Sb species through large-core pseudopotentials Sb ECP46MDF plus (4s4p)/[2s2p] valence BSs \cite{Sb_pseudo}.
The outer 14 O ligands associated with the four adjacent SbO$_6$ octahedra were described through minimal
all-electron ANO BSs \cite{Pierloot1995}.
Large-core pseudopotentials were considered for the 24 Li nearby cations \cite{Fuentealba_Na}.

\noindent
{\it RbCeO$_2$.\,}
We used ECP28MWB quasirelativistic pseudopotentials \cite{Dolg} and Gaussian ANO valence BSs \cite{
Cao_1,Cao_2} for the central Ce species.
All-electron BSs of quintuple-$\zeta$ quality were utilized for the two bridging ligands \cite{Dunning}
and of triple-$\zeta$ quality for the remaining eight O anions \cite{Dunning} of the two octahedra of
the reference magnetic unit.
For the eight Ce neighbors, we employed large-core quasirelativistic pseudopotentials (ECP47MWB)
\cite{Dolg1989,Dolg1993}.
Large-core pseudopotentials were also considered for the 18 Rb nearby cations \cite{Szentpaly,Fuentealba_1983}.
%

\begin{figure}[b]
\centering
\includegraphics[width=0.95\linewidth]{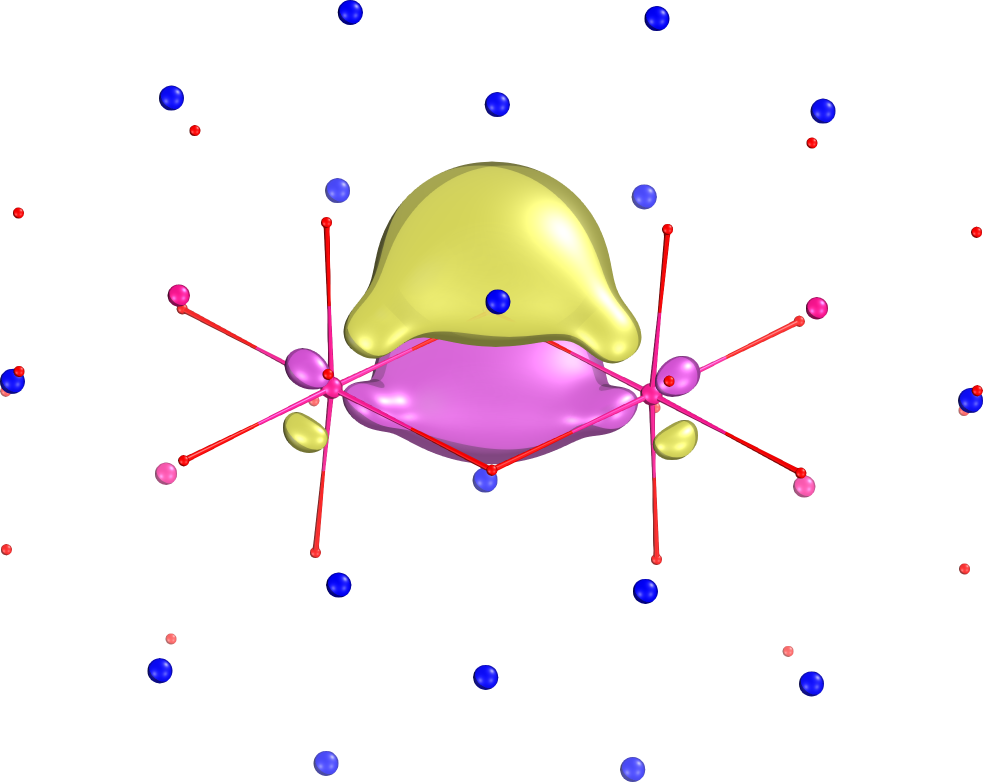}
\caption{
Localized O 2$p$ orbital in Na$_2$IrO$_3$, plot with 90\% of the electron density
within the contour.
}
\label{O2p_Ir213}
\end{figure}

\subsection{Orbital basis for computing exchange contributions}

The analysis of exchange contributions was carried out in terms of localized central-unit orbitals
obtained through Pipek-Mezey localization \cite{PM_method}.
The single-configuration (SC) wavefunctions were constructed using appropriate restrictions for the
occupations of the localized orbitals (LOs), such that intersite excitations are excluded.
From orbital composition analysis through Mulliken partition \cite{Lu_book,Lu_2024}, the tails at
adjacent sites of the magnetic LOs are
$\lessapprox\!1\%$ in RbCeO$_2$,
$\lessapprox\!3\%$ in Li$_3$Co$_2$SbO$_6$,
$\approx\!6\%$ in RuCl$_3$, and
$\approx\!10\%$ in Na$_2$IrO$_3$.
Illustrative LO plots are provided for Li$_3$Co$_2$SbO$_6$ and Na$_2$IrO$_3$ in Figs.\;S1-S3;
the visualization program IboView \cite{knizia2013} was employed.

\bibliography{refs_jun11}